\documentclass[prl,aps,twocolumn]{revtex4}
\usepackage{graphicx} 
\usepackage[usenames]{color}
\usepackage{amsmath,amssymb}
\usepackage{gensymb}
\usepackage{natbib}
\usepackage{xcolor}
\def\strutdepth{\dp\strutbox}
\def\nw#1{\strut\vadjust{\kern-\strutdepth\vtop to0pt{\vss\hbox to\hsize
{\hskip\hsize\hskip5pt$\leftarrow$\hss\strut}}}{\em #1}}
\usepackage{rotating}

\begin{document}

\title{Network topology in soft gels: hardening and softening materials}
\author{Mehdi Bouzid}
\author{Emanuela Del Gado}
\email{emanuela.del.gado@georgetown.edu}
\affiliation{Department of Physics and Institute for Soft Matter Synthesis and Metrology, Georgetown University, 20057 Washington DC, USA}


\begin{abstract}

The structural complexity of soft gels is at the origin of a versatile mechanical response that allows for large deformations, controlled elastic recovery and toughness in the same material. A limit to exploiting the potential of such materials is the insufficient fundamental understanding of the microstructural origin of the bulk mechanical properties. Here we investigate the role of the network topology in a model gel through 3D numerical simulations. Our study links the topology of the network organization in space to its non-linear rheological response preceding yielding and damage: our analysis elucidates how the network connectivity alone could be used to modify the gel mechanics at large strains, from strain-softening to hardening and even to a brittle response. These findings provide new insight for smart material design and for understanding the non-trivial mechanical response of a potentially wide range of technologically relevant materials. 

\end{abstract}

\maketitle

\section{Introduction}
Gels form through phase separation, hierarchical aggregation and self-assembly of soft condensed matter (proteins, colloids or polymers), in many cases made of nano- to micro- particles or agglomerates in suspensions that aggregate into poorly connected and weakly elastic solids, which are very common and ubiquitous in nature \cite{trappe_nature2001,lieleg_nmat2011,helgeson2014homogeneous,conrad-jor2010}. Soft gels are extensively used to improve texture and mechanics in diverse industrial products, where they provide texture, softness and stability. Because of their mechanical versatility and abundance in biocompatible compounds, they are ubiquitous in personal care, foods, and biomedical engineering \cite{gao2015microdynamics,gaharwar2014shear,storm-nature2005}. While such materials are typically highly deformable, under shear they can harden and then fail abruptly, similar to much harder solids: accumulating microscopic damage leads to the nucleation and growth of macroscopic fractures \cite{leocmach2014creep,saint2017predicting,keshawarz2017non,zhang2016fiber}. Especially in protein or biopolymer based gels, their network structure can be sufficiently deformed under shear to put part of the network branches directly under tension, which leads to a strain-hardening response due to their cohesive strength \cite{storm-nature2005,licup2015stress,tabatabai2015rheology,weigandt2009structure}. Under load, eventually the branches can break and initiate a microscopic damage accumulation also associated to creep and delayed crack growth, strongly affected by the load or by the deformation rate and therefore pointing to the role played by the relaxation and redistribution of stresses through the network structure \cite{perge2014time,landrum2016delayed,calzolari2017interplay,Lidon2017}. That is,  the complex and strongly non-linear mechanics of these materials is strongly affected by the network {\it topology}, i.e., the way its connectivity and morphology are organized in space, because it determines how stress can be transferred and redistributed through the material.

Due to the interplay between their microstructure and an imposed deformation, the properties and texture of soft gels can be modified by stretching, flowing or squeezing. Unlike hard materials, these changes can occur over the timescales typical of the material's use or processing in a wide range of applications, and hence determine or compromise their function. Moreover, gel properties can even evolve autonomously in time, due to relaxation of internal stresses frozen into the microstructure during solidification \cite{perge2014time, ferrero2014relaxation, bouzid2017elastically, colombo2014stress}. Hence controlling, and being able to design, the microstructure-process or microstructure-rheology interplay in this class of materials is essential to achieve smart rheology and mechanics that can be finely controlled and adjusted on the fly, such as in soft inks for 3D printing technologies \cite{truby2016printing}. 

A limit to exploiting or expanding the potential of such materials is the insufficient fundamental understanding of the microstructural origin of such diverse combination of mechanical characteristics. Significant attention and efforts have been so far devoted to the diversity of the building blocks of soft gels (particle or droplets of different shape and sizes, rigid or flexible fibers and polymer chains, hierarchically organized agglomerates or bundles ...) and the wide range of possible effective interactions that drive their self-assembly \cite{larson1999structure}. The role of the network topology, instead, remains mainly unexplored and studies that are able to address its complexity and heterogeneity are only nascent \cite{hsiao2012role, wyart-prl2008, valadez2013dynamical, colombo_prl2013, zhong2016quantifying} because of the challenges in extracting relevant information in experiments and a fundamental lack of models and theoretical approaches that can link it to the mechanical response.   

The scope of this work is to attempt to fill this gap, by investigating specifically the link between the connectivity and its spatial organization in the network structure, and the non-linear mechanical response of model soft gels. We use $3D$ numerical simulations of a model particle gel in which we can suitably, and specifically, tune the gel network topology and that we can subject to different types of mechanical tests. In particular, the network structure is made of branches connected by nodes (or crosslinks), whose density in the gel can be simply changed by changing the total particle density or solid volume fraction. Just varying the network topology (without changing the gel components) dramatically change the non-linear response to a shear deformation: very sparsely connected networks favor the presence of soft modes associated to softening, followed by stress localization leading to strain hardening. More densely connected networks with small and homogeneous pores, instead, favor the stresses to distribute uniformly under an applied load, favoring nearly simultaneous breakage of many connections, with a stronger tendency for a brittle response. In contrast, a sparsely connected network more prone to stress localization may entail a more complex crack growth dynamics, and hence a more ductile behavior. Our results suggest that creep, hardening and different fracture modes (brittle vs ductile) could be directly encoded and designed in a soft gel by specifying the network architecture.


\section{Model and numerical simulations}
\label{model}
We use a minimal particle based model that incorporates two basic features of soft gels that are quite generally observed: (i) inter-particle interactions and aggregation kinetics limit the coordination number on certain lengthscales (i.e., at the level of particle-particle contacts or of agglomerates or mesostructures), so that particles assemble into open network-like structures; (ii) strands and nodes of the network possess a certain bending rigidity that provides the network with mechanical stability in spite of locally low degree of coordination~\cite{solomon,hsiao2014model,pantina2005elasticity,dinsmore_prl2006,valadez2013dynamical}. In spite of its simplicity, our model captures important physical features of real colloidal gels and can be used as a prototypical soft amorphous solid.  Starting from this elementary information, we have used anisotropic interactions to introduce local rigidity and stabilize self-assembled thin open structures at low volume fraction \cite{colombo_prl2013,colombo2014self}, in the same spirit as recent works~\cite{delgado2007length,saw2009structural}. Our system consists of $N$ identical particles with position vectors $\{\textbf{r}_i\}\,,i=1\ldots N$,
interacting via the potential energy
\begin{equation}\label{equ:poten}
U(\textbf{r}_1,\ldots,\textbf{r}_N) = \epsilon \left[\sum_{i > j} u_2\left(\frac{\textbf{r}_{ij}}{\sigma}\right) +
 \sum_i\sum_{\substack{j>k}}^{j,k\ne i}u_3\left(\frac{\textbf{r}_{ij}}{d},\frac{\textbf{r}_{ik}}{d}\right)\right]\,,
\end{equation}
where $\textbf{r}_{ij}=\textbf{r}_j-\textbf{r}_i$, $\epsilon$ sets the energy scale and $d$ represents
the particle diameter. Typical values for a colloidal system are $\sigma = 10-100\,\rm{nm}$ and
$\epsilon = 1-100\,k_B T_{\rm r}$, $k_B$ being the Boltzmann constant and $T_{\rm r}$ the room
temperature~\cite{trappe_nature2001,luca_faraday2003,laurati_jor2011}.~The two-body term $u_2$, for particles separated by a distance $r$ (in units of $d$) consists of a repulsive core complemented by a narrow attractive well:
\begin{equation}\label{equ:u2}
  u_2(\textbf{r})=A\left(a\,r^{-18}-r^{-16}\right)\,.
\end{equation}
The three-body term $u_3$ limits the coordination number and confers angular rigidity to the
inter-particle bonds $\textbf{r}$ and $\textbf{r'}$ departing from the same particle:
\begin{equation}
  u_3(\textbf{r},\textbf{r}') = B\,\Lambda(r) \Lambda(r')\,
  \exp\left[-\left(\frac{\textbf{r}\cdot\textbf{r}'}{rr'}-\cos\bar{\theta}\right)^2/w^2\right].
\end{equation}
The range of the three-body interaction is equal to two particle diameters, as ensured by the radial
modulation
\begin{equation}
\Lambda(r)=
\begin{cases}
r^{-10} \left[1-(r/2)^{10}\right]^2 & r<2\,, \\
0 & r\ge 2\,.
\end{cases}
\end{equation} 
The potential energy~\eqref{equ:poten} depends parametrically on the dimensionless quantities $A$, $a$, $B$, $\bar{\theta}$, $w$. We have chosen these parameters such that for $k_B T \sim 10^{-1}\epsilon$ the particles start to self-assemble into a persistent particle network. The data here discussed refer to $A=6.27$, $a=0.85$, $B=67.27$, $\bar{\theta}=65^\circ$, $w=0.30$, one convenient choice to realize this condition.This model has been used to perform a spatially resolved analysis of cooperative dynamics in colloidal gel networks, of their aging and mechanical response ~\cite{colombo_prl2013, colombo2014self,colombo2014stress,bouzid2017elastically}. The network structure is characterized by the cohexistence of poorly connected regions, where major structural rearrangements take place and densely connected domains, where internal stresses tend to concentrate. The structural heterogenity, the long range spatial correlations underlying the cooperative dynamics and the heterogeneous distribution of internal stresses in our model correspond to fundamental physical characteristics of (colloidal) gel networks that have a major role in their complex mechanical response \cite{maccarrone}. 

\subsection{Sample preparation}
We start from gel configurations prepared at $k_B T/\epsilon = 5\cdot 10^{-2}$, whose structure and relaxation dynamics at rest have been already extensively studied ~\cite{colombo_prl2013, colombo2014self}. We have previously determined that in such gels the breaking of bonds is not only determined by the strength of the interaction energy and is affected by the network topology: bond breaking preferentially occurs at the network {\it nodes} (or cross-links) since this is where tensile stresses tend to accumulate \cite{colombo_prl2013}. Nevertheless, for $k_{B}T / \epsilon \ge 10^{-3}$ the effect of the network topology is limited, since bond breaking is still significantly affected by thermal fluctuations \cite{bouzid2017elastically}. Being here specifically interested in understanding how a different spatial organization of the network and of its connectivity can determine a radically (or not) difference in the gel mechanics, we choose to focus on the case $k_B T/\epsilon \simeq 0$ (i.e, the limit of strong attractive interactions), so that we can better isolate the microscopic processes due to the interplay of the network topology with the imposed deformation. 

To this aim, we quench each gel configuration initially prepared at $k_B T/\epsilon=5\cdot10^{-2}$ down to $k_B T/\epsilon \simeq 0$ by using the Langevin dynamics:
\begin{equation}\label{equ:equmotion}
  m\frac{d^2{\textbf{r}}_i}{dt^2} = -\xi\frac{d{\textbf{r}}_i}{dt} - \nabla_{\textbf{r}_i}U\,,
\end{equation}
where $m$ is the particle mass and $\xi$ the coefficient of friction, until the kinetic energy drops to a negligible fraction (less than $10^{-10}$) of its initial value \cite{colombo2014stress,bouzid2017elastically}. For all simulations discussed here we have used $m/\xi=1.0 \tau^{*}$, where $\tau^{*} = \sqrt{md^{2}/\epsilon}$ is the unit time defined by the interactions energy $\epsilon$, the particle diameter $d$ and its mass, and the integration time step is $5\cdot 10^{-3} \tau^{*}$. Having reduced the kinetic energy of the initial samples, the resulting configuration is a local minimum of the potential energy, or \emph{inherent structure}~\cite{stillinger-weber} and we obtain mechanically stable initial samples with $k_B T/\epsilon \simeq 0$ that we can use for rheological tests. 

\subsection{Step shear deformation and imposed shear rate}
On each of the samples prepared following the procedure just described, we perform a series of incremental strain steps in simple shear geometry \cite{colombo2014stress}. In each step we increase the cumulative shear strain by a quantity $\delta\gamma$ by first applying an instantaneous affine deformation $\Gamma_{\delta\gamma}$, corresponding to simple shear in the $xy$ plane, to all particles:
\begin{equation}\label{equ:ss1}
  \textbf{r}_i' = \Gamma_{\delta\gamma}\textbf{r}_i = 
  \begin{pmatrix}
    1 & \delta\gamma & 0 \\
    0 & 1 & 0 \\
    0 & 0 & 1
  \end{pmatrix}
  \textbf{r}_i\,
\end{equation}
The Lees-Edwards boundary conditions are updated as well, to comply with the increase in the cumulative strain \cite{frenkel2002understanding}. The configuration $\{\textbf{r}_i'\}$ is no longer a minimum of the potential energy (being the material amorphous) \cite{alexander}, and the deformation step induces unbalanced internal forces. Hence we relax the affinely deformed configuration by letting the system free to evolve in time while keeping the global strain constant:
\begin{equation}\label{equ:ss2}
  \textbf{r}_i'' = \mathcal{T}_{\delta t}\textbf{r}_i'\,.
\end{equation}
where $\mathcal{T}_{\delta t}$ is the time evolution operator for the Langevin dynamics~\eqref{equ:equmotion} and a specified time interval $\delta t$. After $n$ steps, the cumulative strain is $\gamma_n=n\,\delta\gamma$ and the gel configuration is
\begin{equation}
  \textbf{r}_{i,n} = (\mathcal{T}_{\delta t}\Gamma_{\delta\gamma})^n\,\textbf{r}_{i,0}\,,
\end{equation}
where $\{\textbf{r}_{i,0}\}$ denotes the configuration of the starting inherent structure. 

The procedure just outlined is similar to the \emph{athermal quasistatic} (AQS) approach extensively used to investigate the deformation behavior of amorphous solids~\cite{tanguy2002continuum,maloney2006amorphous,fiocco-pre2013}, the main difference being that, instead of using an energy minimization algorithm after each affine deformation step, here we follow the natural dynamics of the system (with viscous energy dissipation) for a prescribed time interval $\delta t$. Such approach, therefore, allows us to define a finite shear rate $\dot{\gamma} = \delta\gamma/\delta t$ for the deformation we apply. 

Disregarding effects due to the particle inertia, the microscopic dynamics~\eqref{equ:equmotion} introduce a natural time scale $\tau_0=\xi d^2/\epsilon$, corresponding to the time it takes a particle subjected to a typical force $\epsilon/\sigma$ to move a distance equal to its size. In all simulations discussed here we fix the elementary strain increment $\delta\gamma=10^{-2}$, and choose the relaxation interval $\delta t$ to obtain a shear rate $\dot{\gamma}_s=10^{-5} \tau_{0}^{-1}$. Indicatively, if we consider a typical aqueous solution of colloidal particles with a diameter $d \approx 100$ nm and an interaction energy $\epsilon\approx 10k_B T $~\cite{petekidis_softmatter2011} the characteristic time is $\tau_0\approx 10^{-4}$ s; in such a system the rate we have investigated here would correspond to 0.1 $\rm{s}^{-1}$.

We compute the global stress tensor $\sigma_{\alpha\beta}$ using the standard virial equation~\cite{thompson2009general} at the end of each deformation step:
\begin{equation}
  \sigma_{\alpha\beta} = \frac{1}{V} \sum_{i=1}^N \frac{\partial U}{\partial r_i^\alpha}\,r_i^\beta\,,
\end{equation}
in which $V$ is the volume of the system and $\alpha,\beta$ stand for the cartesian components $\{x,y,z\}$. In the following we use $\sigma$ to indicate the shear component $\sigma_{xy}$ in the mechanical tests.
Since the velocities of the particles are small ($v_i\lesssim 10^{-5} d/\tau_0$) in the calculation of the stress tensor we ignore the kinetic term $mv_i^{\alpha}v_i^\beta$ and any contribution due to the viscous forces appearing in Eq.~\eqref{equ:equmotion}. 

With the approach just described we obtain the load curve of the material at different shear rates and analyse the underlying microscopic processes, in terms of local stresses and strains and of structural modifications of the gel.

\subsection{Oscillatory rheology tests}

For each of the gel configurations, in addition to the transient step-strain tests, we investigate the linear and non-linear viscoelastic properties by measuring the frequency and the strain dependence of the first-harmonic storage $G'$ and loss modulus $G''$ \cite{larson1999structure,mewis2012colloidal}. The computational scheme in this case consists in imposing an oscillatory shear strain on the system, i.e., the shear strain is modulated periodically according as $\gamma(t)=\gamma \sin(\omega t)$. The equation of motion is 
\begin{equation}
m\frac{d^2\bold{r}_i}{dt^2}=-\nabla_{\bold{r_i}}\mathcal{U}-\eta_f\left(\frac{d\bold{r}_i}{dt}-\dot\gamma(t)y_i\bold{e_x}\right)
\label{shear}
\end{equation}
and we use Lees-Edwards boundary conditions as in the step-strain deformation test. 
By monitoring the shear stress response of the material $\sigma(t)$ over time, we can extract the viscoelastic moduli. The storage and the loss modulus can be computed from the stress response with the following expressions:
\begin{align}
G'(\omega)=\mathcal{R}e\left(\frac{\bar{\sigma}(\omega)}{\bar{\gamma}(\omega)}\right) \label{RE}\\
G''(\omega)=\mathcal{I}m\left(\frac{\bar{\sigma}(\omega)}{\bar{\gamma}(\omega)}\right) \label{IM}
\end{align}
Where $\bar{\sigma}$ and $\bar{\gamma}$ are the Fourier transforms of respectively the shear stress and the strain. For a fixed strain amplitude $\gamma$ in the linear response regime, we vary the frequency $\omega$ to explore the linear viscoelastic spectrum. For a fixed frequency $\omega\tau_0=10^{-3}$, for which the material response is dominated by its elastic component for all volume fractions considered here, we have then varied the strain amplitude $\gamma$ from $1\%$, which is in the linear response regime, to $400\%$, i.e., in the non-linear regime preceding yielding. 

All simulations were performed using the LAMMPS molecular dynamics source code~\cite{plimpton1995fast}, which we have suitably extended to include the interaction~\eqref{equ:poten}. The gel consists of $N=1.08\cdot10^5$ particles in a cubic simulation boxes with linear size $L=102.6d, 90d, 81.44d, 71.44d$ corresponding to approximate volume fractions $\phi \simeq 0.05, 0.075, 0.10, 0.15$. 

\section{Results and discussion}
\label{results}
The mechanical tests just described help us characterize the linear and non-linear rheological response of the gels. Here we combine such information with a microscopic analysis on the different network topologies corresponding to different volume fractions in the model.
\begin{figure}
\includegraphics[width=0.99\linewidth]{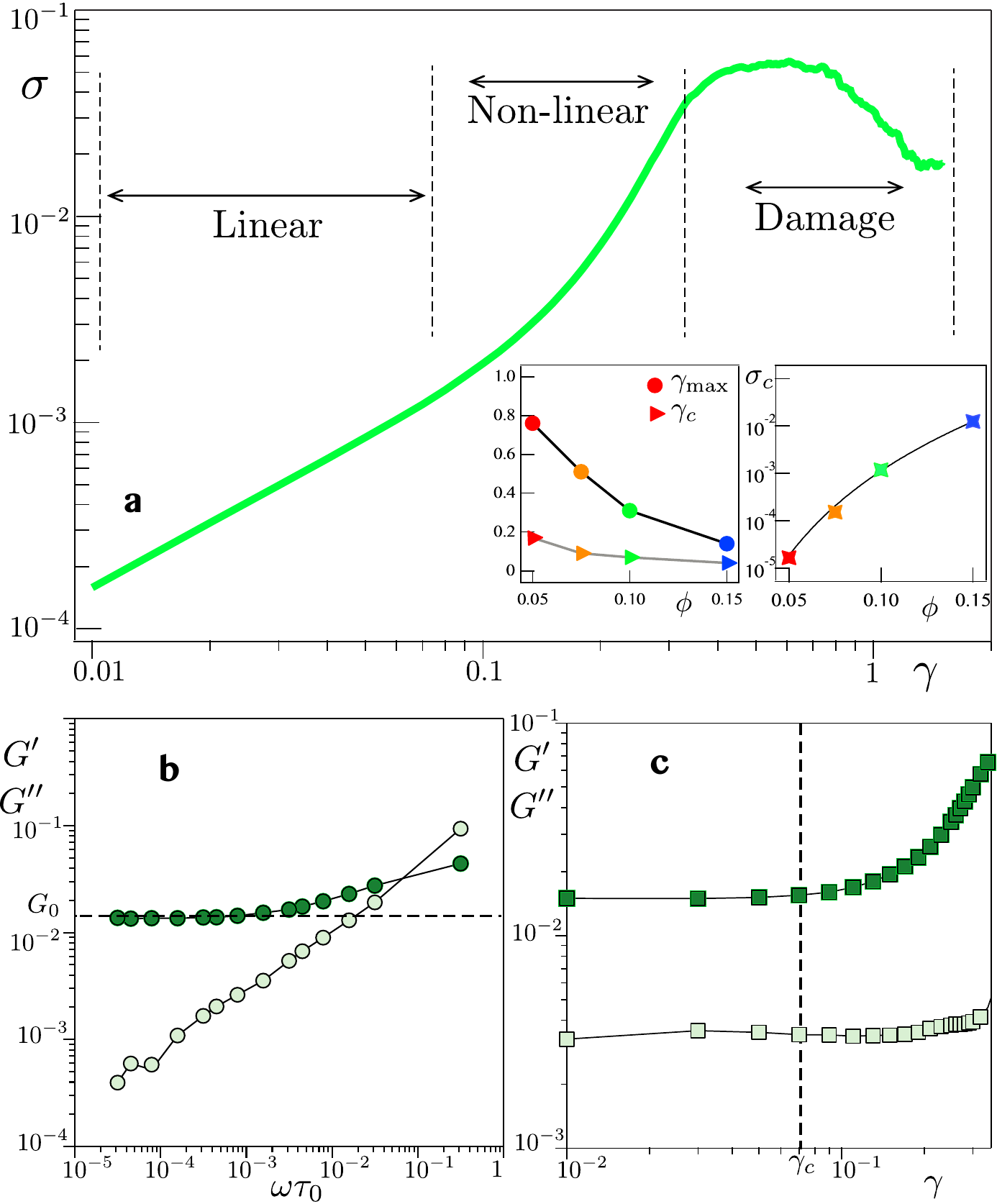}
\caption{\label{Fig1} (a) Load curve: the total shear stress $\sigma$ as a function of the applied strain, $\gamma$ for a gel of volume fraction $\phi=10\%$. The black vertical lines represent the different mechanical regimes, as discussed in the text. Insets show the critical strain $\gamma_c$ and shear stress $\sigma_c$ (corresponding to the onset of the strain stiffening), as well as the maximum strain $\gamma_{\rm max}$ (corresponding to the end of this regime before yielding starts), as a function of the volume fraction. (b) Linear oscillatory response: storage and loss moduli, $G'$ (dark circles)  and $G''$ (light circles) as a function of the normalized frequency $\omega$.  (c) Strain sweep:  $G'$ (dark squares)  and $G''$ (light squares) as a function of the strain amplitude $\gamma$ obtained at a fixed frequency $\omega\tau_0=10^{-3}$, displaying strain stiffening above a critical strain amplitude $\gamma_c\sim 0.07$.}
\end{figure}
Fig.~\ref{Fig1} provides an overview of the gel mechanics obtained from the different tests performed at $\phi =10\%$: the load curve (a) of the material obtained from the step-strain simulations indicates a linear elastic response at small deformations, followed by a non-linear regime until the material starts to yield. At the low shear rate considered here, the yielding is accompanied by a significant strain localization and damage of the gel structure, as described in \cite{colombo2014stress}. The linear oscillatory response in a frequency sweep (b) shows that $G' \gg G''$ for $\omega < 10^{-2} \tau^{-1}_{0}$ and reaches a constant value $G_0$ in the limit of vanishing frequency. Varying the strain amplitude in the oscillatory tests, we obtain the strain-sweep (c) which shows how above a critical strain amplitude the material is strain-stiffening, with $G'$ significantly growing while we do not detect any significant increase in the dissipative response. 

Having defined $\sigma_{c}$ as the critical stress above which the system exhibits a strain stiffening, we identify the corresponding critical strain $\gamma_c$ in the stress strain curve. From the material response $\sigma (\gamma)$, we compute the differential modulus $K\equiv \frac{\partial\sigma}{\partial\gamma}$. We identify the end of the stiffening regime (and the beginning of the yielding) in terms of the strain $\gamma_{max}$, defined as the maximum of the differential modulus $K$. The dependence of $\gamma_{c}$, $\sigma_c$ and $\gamma_{max}$ on the volume fraction, and hence on the gel topology, is shown in the inset of Fig.\ref{Fig1} (a).
 
\begin{figure}
\includegraphics[width=1\linewidth]{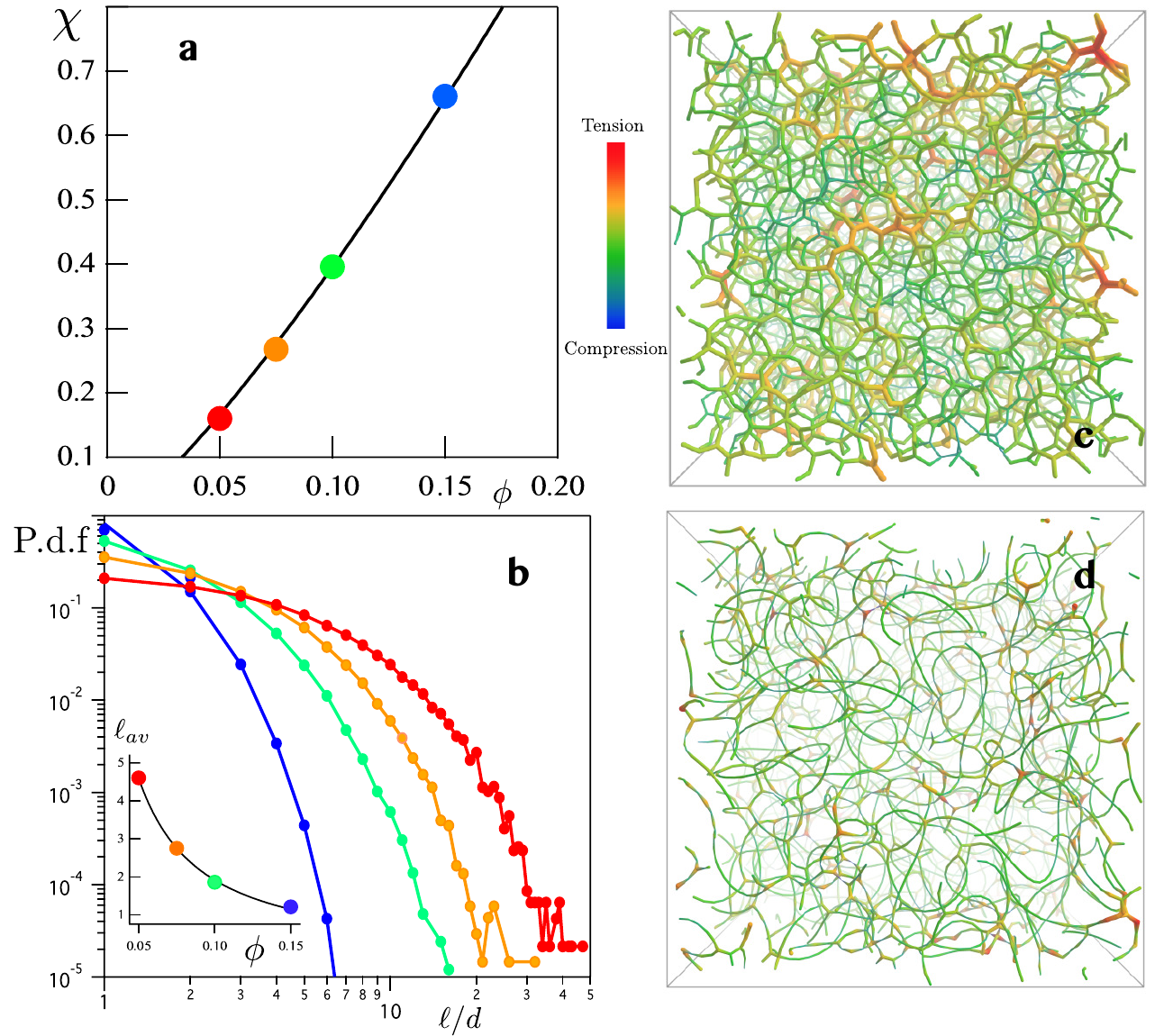}
\caption{ (a) The fraction $\chi$ of the total number of bonds that participate to network nodes (defined by particles with a coordination number 3) as a function of the volume fraction for the initial configurations after solidification. (b) The length distribution of the gel branches between two nodes for different volume fraction, the inset shows the average chain length as a function of the volume fraction. Snapshots of the gel network extracted from the simulations at a volume fraction $\phi\simeq 15\%$ (c) and $\phi\simeq 5\%$ (d). Each bond is represented by a segment, when the distance $d_{ij}$ between two particles $i$ and $j$ is $d_{ij}\leq 1.3d$. The color code shows the value of local tensile or compressive stresses, while the thickness is proportional to the stress amplitude (see Supplementary Information).}
\label{Fig2} 
\vspace{-0.2cm}
\end{figure}

Varying the volume fraction in the model allows us to change the topology of the gel network in a very specific way. As shown in Fig. \ref{Fig2} (a), having followed the same preparation protocol for all samples, increasing the volume fraction produces an increase in the fraction $\chi$ of bonds that are involved in the network {\it nodes}, defined as particles with coordination number $3$. Hence, upon increasing the volume fraction, the gels are more densely connected with typical mesh sizes that are progressively reduced. Fig.\ref{Fig2}(b) displays the distribution of the length of the gel branches (i.e., chains made of particles with coordination number 2) connecting two nodes of the network, measured along the network. Such distributions tend to follow an exponential law (as found in the studies of the gel self-assembly in this model \cite{colombo2014self}) and their width and mean value increases with decreasing the volume fraction (see also inset). The change in the gel morphology is illustrated through two snapshots from the simulations (Figs.\ref{Fig2}(c) and (d)) that show only the network links for clarity and correspond respectively to volume fractions $0.05$ and $0.15$. The connections of the gel networks have been colored according to the local stresses (tensile or compressive, see also Supplementary Information), and their thickness is proportional to the stress magnitude. The snapshots highlight how in the more sparse network higher stresses are concentrated in very few nodes, while in the more densely connected network they are more distributed through the network structure.  

We use the mechanical and topological characterization described so far to establish a link between the gel response to an imposed deformation and the gel topology, by analyzing samples prepared at different volume fraction $\phi$. By varying the volume fraction, in addition to the obvious increase of the low frequency modulus of the gels with $\phi$, we observe striking qualitative changes in the non-linear regime of the load curve and of the strain-sweeps. 



\subsection{Softening and hardening}
\begin{figure}
\includegraphics[width=1\linewidth]{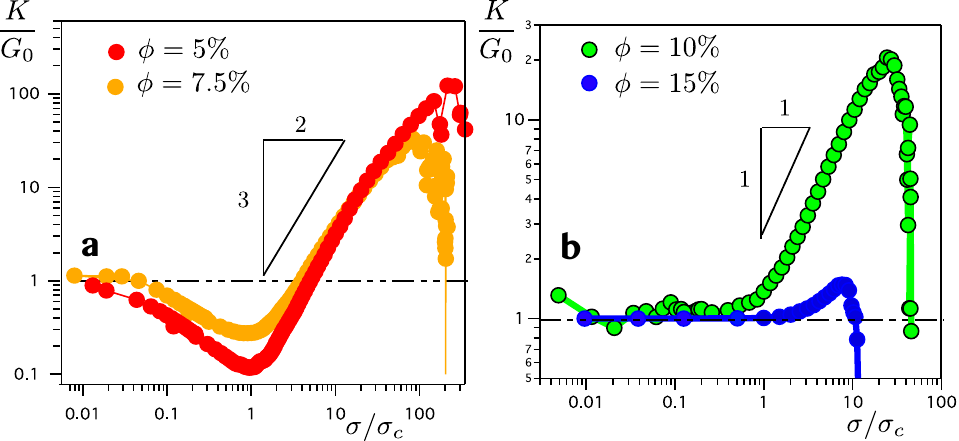}
\caption{(a) The differential modulus $K$ normalized by the elastic modulus $G_0$ as a function of the normalized shear stress $\sigma/\sigma_c$, for $\phi=5\%$ and $\phi=7.5\%$ and for $\phi=10\%$ and  $\phi=15\%$ (b).}
\label{Fig3} 
\end{figure}

When rescaling the differential modulus $K$ by the low frequency elastic modulus $G_0$ and the stress measured by $\sigma_{c}$, we can directly compare the non-linear response of gels at different volume fractions, independently of their different stiffness in the linear response regime. Fig.\ref{Fig3} (a) shows how, for relatively small volume fraction ($\phi=5\%$ and $\phi=7.5\%$), the gels exhibit a linear elastic response at small deformations (the initial flat part of $K/G_{0}$ as a function of $\sigma/\sigma_{c}$), followed by a pronounced softening before entering a strain-hardening regime at larger deformations \cite{weigandt2009structure, tabatabai2015rheology,kim2014structural,xu2011strain}. The strain hardening is characterized by a power law scaling $K/G_0\sim(\sigma/\sigma_c)^{\alpha}$, similar to the one found in semiflexible polymer networks \cite{jamney-natmat2007, broedersz2014modeling} with an exponent $\alpha=3/2$, at $\phi=0.05, 0.075$. In those systems this type of dependence is ascribed to the entropic contribution to the elastic free energy due to the semiflexibility of the gel branches. The bending stiffness in our model provides semiflexibility to the gel networks in our study \cite{colombo2014self,colombo2014stress} but the response to deformation here is dominated by enthalpic contributions since thermal fluctuations are neglected. 
Nevertheless, the amount of deformation that can be accomodated in the system before overstretching most of the gel branches is strongly affected by the disorder and the topological heterogeneity of the gel network. Hence the similarity of the strain stiffening regime found here in the poorly connected networks with the one typically observed in semiflexible polymer networks could be understood in terms of an entropic-like contribution to the elastic energy emerging at the level of the network from the disorder and the topological heterogeneity, whose origin is again the flexibility of the gel branches. Such picture is consistent with the fact that the stiffening in the soft gels, characterized by the exponent $3/2$, is a purely non-linear elastic regime, where there is no (or hardly any) breaking of existing bonds or formation of new ones. 
At moderate ($\phi=10\%$) volume fraction, the softening disappears and the linear elastic response extends over larger deformations (see inset of Fig.\ref{Fig1}(a)). The following strain-stiffening, while reminiscent of the non-linear response measured at lower volume fractions, has a different (linear) scaling with the shear stress $K/G_0\sim\sigma/\sigma_c$: since the differential modulus $K = \partial \sigma /\partial \gamma$, the fact that it has a linear dependence on $\sigma$ indicates that there is an exponential increase of the shear stress with the strain in the non-linear regime. Such regime is reminiscent of the one found in fiber networks models \cite{licup2015stress, zhang2016fiber} and can be understood since, by increasing the volume fraction and hence the nodes density in the network, the gel branches are more rigid (being shorter and subjected to stronger topological constraints).   
We note that whereas no significant changes in the network connectivity and topology (i.e. through breaking and reforming of bonds) underlies the strain softening and stiffening at the lowest volume fractions, upon increasing the volume fraction ($\phi =10\%$) strain induced bond formation starts already at very low strains in the stiffening regime and bond-breaking starts to take place during the stiffening regime as well (once a certain fraction of the gel branches has been straightened out), suggesting that a feedback between the breaking of bonds in weaker areas of the material and formation of new ones may set in and lead to a strain hardening of the material \cite{colombo2014stress}.

Finally, at the highest volume fraction investigated ($\phi=15\%$), where the fraction of bonds that participates to network nodes is the highest, we hardly detect any non-linear response before the material is damaged and eventually yields (see inset Fig.\ref{Fig1}(a)), where $\gamma_{c} \simeq \gamma_{max}$ at this volume fraction. The inset of Fig.\ref{Fig1}(a) also shows that $\sigma_{c}$, the critical load beyond which the response is non-linear, progressively decreases with increasing the volume fraction. Overall, the denser gels (which are also more homogeneously connected) tend to have a rather brittle response, where within a small increase in the strain the gel fails or yields, contrasting with the extremely ductile behavior found in the less connected gels, which can deform extensively before yielding.

\begin{figure*}[t!]
\includegraphics[width=1\linewidth]{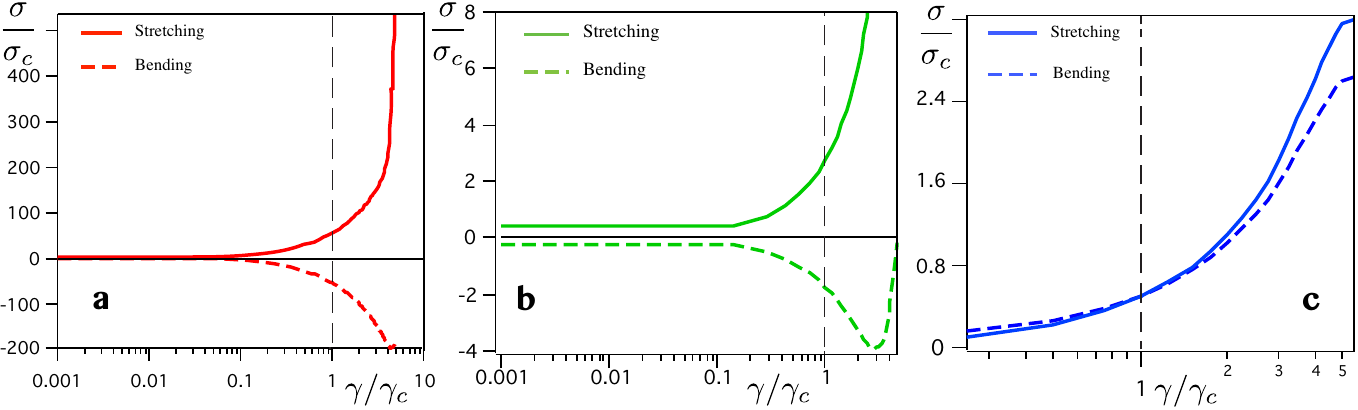}
\caption{The different shear stress contributions:  stretching (solid lines) and bending (dashed) for volume fractions that correspond respectively to $\phi\sim5\%$ (a), $\phi\sim10\%$ (b) and $\phi\sim15\%$ (c). }
\label{Fig4} 
\end{figure*}

Further insight is gained when we separate the contributions to the shear stress coming from different part of the effective interactions employed in the model. The 2-body part of the interaction potential described above, in fact, is mainly responsible for the stretching of the gel branches, whereas the 3-body one introduces a bending rigidity to branches and nodes of the gel. At low volume fractions (Fig.\ref{Fig4} (a)), where the gel network is sparse and gel branches longer, the onset of the non-linear regime is characterized by equal amounts of stretching and bending, with the two contributions being opposite in sign. That is, the stretching and bending stresses compete while respectively resisting and favoring the deformation. By monitoring the fraction of bonds that is broken in the network upon increasing the strain, we can clearly recognize that the softening detected here is not associated to a net decrease in the network connectivity, since there is no broken or newly formed bonds at the corresponding strains. Such findings suggest instead that, since at these low volume fractions the gel branches can be initially bent (a left-over of the thermal fluctuations initially present, due to the limited topological constraints), the imposed deformation ends up unbending them (or stretching them out) and releasing bending stresses. 

Upon increasing the network connectivity, the softening disappears and the stiffening dominates  (Fig.\ref{Fig4}(b)): for such gels, the stretching and bending contributions are still opposite in sign (and therefore competing with each other) but the stretching controls the overall mechanical response of the gel \cite{feng2015alignment,feng2016nonlinear}, consistent with a network whose branches can be more stretched out in the initial configuration but bending stresses cannot be easily released, due to the increased topological constraints, so that the mechanism proposed in the previous case does not prevail. On the other hand, we note that in this case the gels can sustain larger strains before yielding starts (see the corresponding value of $\gamma_c$ in the inset of Fig.\ref{Fig1} (a)). Such behavior suggests that, although the topological constraints have increased, the network connectivity is still sufficiently sparse and heterogeneous to accomodate large deformations by stretching out the regions of the gel where the connectivity is lower \cite{colombo2014stress,feng2015alignment}. This scenario is consistent with a response that is prevalently strain-stiffening and with a localization of the stresses in the stretched out parts of the gels, where local tension has increased and breaking is more likely to occur.  

Finally, by increasing the volume fraction and hence the network connectivity further, both stretching and bending contribution to the shear stress concur to resist to the deformation and remain very similar in magnitude (Fig.\ref{Fig4} (c)): this is the case in which there is a rather abrupt transition from the linear regime to the yielding, with hardly any non-linearity in between, and corresponds to a gel that is highly and more homogeneously connected. The data suggest that the stress redistribution under deformation is not dominated by a strong localization, as in the previous case, and is characterized, instead, by a more homogeneous repartition of the stresses through the network structure, certainly helped by a more homogeneously distributed connectivity. This type of material is able to sustain higher stresses at the onset of the non-linear response (see the corresponding value of $\sigma_{c}$ in the inset of Fig.\ref{Fig1}) since it is stiffer, but the deformation it can accomodate before yielding is much more limited and the mechanical failure quite abrupt.

\begin{figure}
\includegraphics[width=1\linewidth]{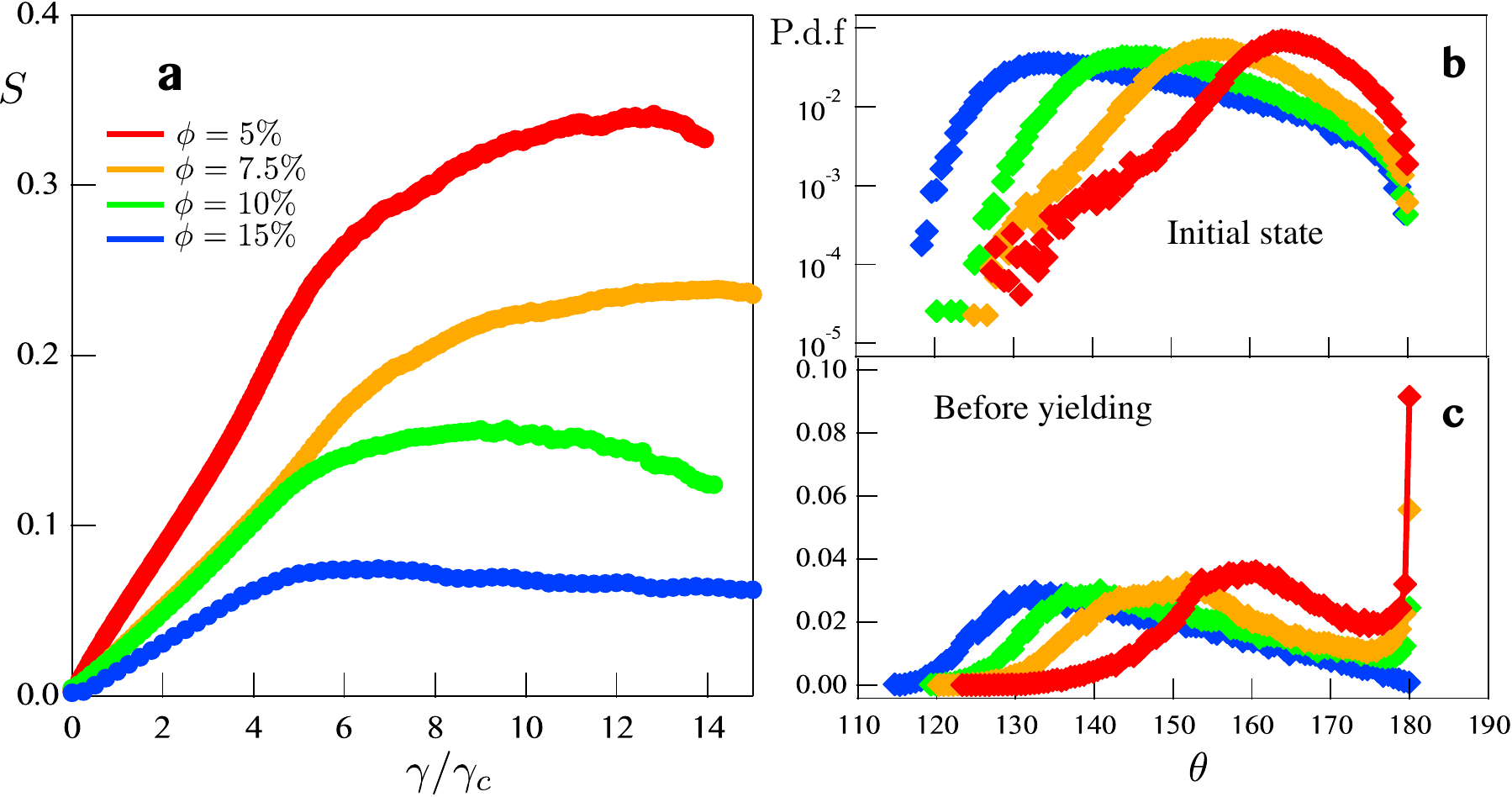}
\caption{(a) The scalar nematic order parameter $S$ as a function of the normalized strain $\gamma/\gamma_c$. (b) Histogram of the bond angles $\theta$  in the initial configuration for gels at different volume fractions and (c) corresponding to shear strains at the end of the strain hardening, right before the yielding.}
\label{Fig5}  
\end{figure}

Analyzing further the morphological changes of the gel networks under deformation in the different cases help us confirm the scenarios just discussed. The topological constraints that change upon changing the volume fraction in our model translate into a stronger or weaker tendency of the gel structure to orient following or not the imposed deformation. The structural anisotropy emerging under deformation in gels and other soft solids has been successfully often used to characterize their response and to obtain more information on the microstructural origin of the rheological response \cite{feng2015alignment, rezakhaniha2012experimental,vermant_jor2009,jamali2017microstructural,boromand2017structural,colombo2014stress,wang2016large}. Here we quantify such tendency in terms of a nematic tensor $\mathbf{Q}_{\alpha\beta}$ obtained from the second moment of the bond orientations given by the unit vector $\bf{n}$ pointing between a particle $i$ and its $j$th bond forming neighbor. $\mathbf{Q}_{\alpha\beta}=\frac{1}{2}\langle3\mathbf{n}_{\alpha}\mathbf{n}_{\beta}-\delta_{\alpha\beta}\rangle$, where $\langle \cdot \rangle$ is the average over all bonds. In order to characterize the strength of the alignment, we compute a scalar order parameter $S$, defined as the largest positive eigenvalue of $\mathbf{Q}$ \cite{chaikin2000principles}. $S$ is bounded between $0$ for a random orientation and $1$ for a system where all the bonds are fully oriented along the same direction. When plotted as a function of the normalized strain $\gamma/\gamma_c$ (Fig.\ref{Fig5} (a)), $S$ highlights the correlation between the degree of anisotropy (or nematic order) in the bond orientation and the distinctive features of the non-linear response of the gels. The value of $S$ increases with increasing the normalized strain and reaches a plateau in the non-linear regime that strongly increases with decreasing the volume fraction, that is, going from gels that are topologically more homogeneous and stiffer to gels that are locally softer and sparse. To combine this global observable with more local information, we compute the distribution of the angles $\theta$ between bonds connecting neighboring particles through the gels at different volume fractions in the initial configurations at rest (Fig.\ref{Fig5} (b)) and in the non-linear regime right before yielding (Fig.\ref{Fig5} (c)). 
In the initial configurations, locally softer gels are characterized by a wider distribution of bond-angles. We emphasize that the more pronounced tails for small angles are consistent with a fraction of the gel branches that are bent and their unbending determines the softening described above. The relative weight of small {\it vs} large angles (bent {\it vs} unbent branches) point to a highly localized nature of the softening regime. In such gels the shape of the bond angle distribution has significantly changed by the end of the stiffening regime, where a set of fully stretched chains appears and grows. The changes in the bond angles are reduced with increasing the volume fraction: eventually for locally stiff and topologically more homogeneous gels there is hardly any change as the system starts yielding and no sign of stretched out chains, consistent with the absence of strain stiffening.  

\section{Conclusion}
\label{conclu}
The emerging picture is that the topology of soft gels can have a major role in dramatically changing their mechanical response, due to a drastically different stress redistribution under deformation that, in this type of materials, can be obtained with relatively small changes of the topology. An important implication is that the local softness of the material should be thought of as topologically controlled (as opposed, for example, to just consider weaker or stronger effective interactions): locally soft gels would be characterized by a sparse stress bearing networks where the soft modes associated to the unbending of the gel branches can dominate the non-linear response at the point of making it significantly strain-softening. Locally stiffer gels can lack such soft modes and exhibit purely stiffening behavior thanks to their still relatively sparse connectivity that allows for stretching and orienting the gel branches along the direction of maximum elongation in shear \cite{colombo2014stress}. Finally, for locally even stiffer gels that are also topologically more homogenous, stress redistribution will also be more homogeneous, bond reorientation is very limited, stretching and bending of the gel structure can equally contribute to resist to the deformation: in this case the response to deformation tends to be more brittle, with a rapid transition from the elastic regime into the yielding and failure \cite{zhang2016fiber}. Parts of this picture are consistent with several experimental observations (e.g., for the softening and hardening regimes in protein and biopolymer gels) \cite{storm-nature2005, storm-nature2005,licup2015stress,tabatabai2015rheology,weigandt2009structure}, while novel experiments able to combine imaging or spectroscopy and rheology could help establish a more stringent assessment of the ideas put forward here \cite{arevalo2015stress,vermant_jor2009,eberle2012flow}. The analysis based on the order parameter $S$ supports the idea that the non-linear mechanical response of soft gels is controlled by topological defects (the network nodes that hinder the alignment to the imposed deformation) and our results suggest that their density and the nature of their spatial correlations are at the origin of the dramatically different behavior of such materials. Overall our findings provide new insight into how to establish the sought-after link between the microstructure and the bulk mechanical properties of soft gel networks, into how to develop a more general condensed matter theoretical framework for their mechanics and, potentially, into how to design non-linear response and long-term evolution of smart soft materials. 

\textit{Acknowledgement:}
The authors thanks Dan Blair, Thibaut Divoux, Pasha Tabatai, Jeff Urbach, Robin Masurel, Peter Olmsted for insightful discussions. 
This work was supported by the Impact Program of the Georgetown Environmental Initiative and Georgetown University.

\section{Supplementary informations}
 \section{Calculation of coarse-grained local stresses}
At each timestep, we characterize the state of stress of a gel configuration by computing the virial stresses as 
$ \sigma_{\alpha\beta}=-\frac{1}{V}\sum\limits_{i}w_{\alpha\beta}^i$, where the Greek subscripts stand for the Cartesian components  $ {x,y,z}$ and $ w_{\alpha\beta}^i$ represents the contribution to the stress tensor of all the interactions involving the particle $ i$ and $ V$ is the total volume of the simulation box \cite{thompson2009general}.
$ w_{\alpha\beta}^i$ contains, for each particle, the contributions of the two-body and the three-body forces evenly distributed among the particles that participate in them:
\begin{equation}
w^i_{\alpha\beta}=-\frac{1}{2}\sum\limits_{n=1}^{N_2}(r_\alpha^iF_\beta^i+r_\alpha^{\prime}F_\beta^{\prime})+\frac{1}{3}\sum\limits_{n=1}^{N_3}(r_\alpha^iF_\beta^i+r_\alpha^{\prime}F_\beta^{\prime}+r_\alpha^{\prime\prime}F_\beta^{\prime\prime})
\end{equation}
The first term on the r.h.s. denotes the contribution of the two-body interaction, where the sum runs over all the $N_2$ pair of interactions that involve the particle $ i$. $ (r^i,F^i)$ and $ (r^{\prime},F^{\prime})$ denote respectively the position and the forces on the two interacting particles. The second term indicates the three-body interactions involving the particle $ i$. 
We consider a coarse-graining volume $\Omega_{cg}$ centered around the point of interest $\bf{r}$ and containing around 9-10 particles on average, and define the local coarse-grained stress based on the per-particle virial contribution as $ \tilde{\sigma}_{\alpha\beta}(\bf{r})=-\sum\limits_{i\in\Omega_{cg}}w_{\alpha\beta}^{i} / \Omega_{cg}$.
For a typical starting configuration of the gel, the local normal stress $ \tilde{\sigma_n}=(\tilde{\sigma}_{xx}+\tilde{\sigma}_{yy}+\tilde{\sigma}_{zz})/3$ reflect the heterogeneity of the structure and tend to be higher around the nodes, due to the topological frustration of the network.
%



\providecommand{\latin}[1]{#1}
\providecommand*\mcitethebibliography{\thebibliography}
\csname @ifundefined\endcsname{endmcitethebibliography}
  {\let\endmcitethebibliography\endthebibliography}{}

\end{document}